\documentclass[12pt]{article}
  
\usepackage{amsmath,amssymb}
\usepackage{bm}
\usepackage{graphicx, color}
\usepackage{wrapfig}
\usepackage{cite}
\begin{document}
\title{
\begin{flushright}
\ \\*[-80pt] 
\begin{minipage}{0.21\linewidth}
\normalsize
APCTP Pre2018-20
 \\*[50pt]
\end{minipage}
\end{flushright}
{\Large \bf 
  CP violation of quarks in  $A_4$ modular  invariance 
\\*[20pt]}}

\author{ 
\centerline{
Hiroshi Okada $^{a}\footnote{E-mail address:hiroshi.okada@apctp.org}$~ and 
~~Morimitsu Tanimoto $^{b}\footnote{E-mail address: tanimoto@muse.sc.niigata-u.ac.jp}$} \\*[5pt]
\centerline{
\begin{minipage}{\linewidth}
\begin{center}
$^a${\it \normalsize
Asia Pacific Center for Theoretical Physics (APCTP) - Headquarters
San 31, Hyoja-dong, Nam-gu, Pohang 790-784, Korea} \\*[5pt]
$^b${\it \normalsize
Department of Physics, Niigata University, Niigata 950-2181, Japan}
\end{center}
\end{minipage}}
\\*[50pt]}

\date{
\centerline{\small \bf Abstract}
\begin{minipage}{0.9\linewidth}
\medskip 
\medskip 
\small 
We discuss the quark mass matrices in the $A_4$  modular symmetry,
where the $A_4$ triplet of Higgs is introduced
for each  up-quark and down-quark sectors, respectively.
The model has six real parameters and two complex parameters
in addition to the modulus $\tau$.
By inputting six quark masses and three CKM mixing angles,
we can predict the CP violation phase $\delta$ and the Jarlskog invariant $J_{CP}$.
The predicted ranges of  $\delta$ and $J_{CP}$   are 
consistent with the observed values.
The absolute value of $V_{ub}$ is smaller than $0.0043$,
while $V_{cb}$ is lager than $0.0436$.
In conclusion, our quark mass matrices with the  $A_4$  modular  symmetry
can reproduce the CKM mixing matrix completely with observed quark masses.
\end{minipage}
}

\begin{titlepage}
\maketitle
\thispagestyle{empty}
\end{titlepage}

\section{Introduction}

The origin of three families of quarks and leptons remains most important problems of Standard model (SM). 
In order to understand the  flavor structure of quarks and leptons,
considerable interests in the discrete flavor symmetry \cite{Altarelli:2010gt,Ishimori:2010au,Ishimori:2012zz,Hernandez:2012ra,King:2013eh,King:2014nza,Tanimoto:2015nfa,King:2017guk,Petcov:2017ggy}
have been developed by the early models of quark masses and mixing angles 
\cite{Pakvasa:1977in,Wilczek:1977uh},
more recently, the large flavor mixing angles of the leptons.

Many models have been proposed by using $S_3$, $A_4$, $S_4$, $A_5$ and other groups with lager orders to explain the large neutrino mixing angles.
Among them, the $A_4$ flavor model is attractive one 
because the $A_4$ group is the minimal one including a triplet 
 irreducible representation, 
which allows for a natural explanation of the  
existence of  three families of leptons 
\cite{Ma:2001dn,Babu:2002dz,Altarelli:2005yp,Altarelli:2005yx,Shimizu:2011xg,
	Kang:2018txu}.
However, variety of models is so wide that it is difficult to obtain clear clues of the $A_4$ flavor symmetry.
Indeed, symmetry breakings are required to reproduce realistic mixing angles
\cite{Petcov:2018snn}.
The effective Lagrangian of a typical flavor model is given by introducing the gauge singlet scalars which are so-called flavons.
Their vacuum expectation values (VEVs) determine the flavor structure of quarks and leptons.
As a consequence, the breaking sector of flavor symmetry typically produces many unknown parameters.

Recently, new  approach to the lepton flavor problem
based on the invariance under the modular group \cite{Feruglio:2017spp}, 
where the model of the finite
modular group  $\Gamma_3 \simeq A_4$ has been presented.
This work  inspired further studies of the modular invariance 
approach to the lepton flavor problem. 
It should be emphasized that there is a significant difference between the 
models based on the $A_4$ modular symmetry and those based on the usual 
non-Abelian discrete $A_4$ flavor symmetry.
Yukawa couplings transform non-trivially under the modular symmetry 
and are written in terms of modular forms which are  
holomorphic functions of a complex parameter,  the modulus  $\tau$.

It is interesting that the modular group includes $S_3$, $A_4$, $S_4$, and $A_5$ as its finite subgroups \cite{deAdelhartToorop:2011re}.
Along the work of the $A_4$ modular group \cite{Feruglio:2017spp}, models of $\Gamma_2 \simeq S_3$ \cite{Kobayashi:2018vbk},
$\Gamma_4 \simeq S_4$ \cite{Penedo:2018nmg} 
and  $\Gamma_5 \simeq A_5$ \cite{Novichkov:2018nkm} have been proposed.
Also numerical discussions of the neutrino flavor mixing have been done
based on  $A_4$ \cite{Criado:2018thu,Kobayashi:2018scp} and $S_4$ \cite{Novichkov:2018ovf}  modular groups respectively.
In particular, the comprehensive analysis of the $A_4$ modular group 
has provided a clear prediction of the neutrino mixing angles and the CP violating phase \cite{Kobayashi:2018scp}.
On the other hand, the $A_4$ modular symmetry has been applied to the $SU(5)$ grand
unified theory of  quarks and leptons  \cite{deAnda:2018ecu},
and also the residual symmetry of the $A_4$ modular symmetry has been investigated \cite{Novichkov:2018yse}.
Furthermore, modular forms for $\Delta(96)$ and $\Delta(384)$ were constructed \cite{Kobayashi:2018bff},
and the extension of the traditional flavor group  is discussed with modular symmetries \cite{Baur:2019kwi}.

In this work, we discuss the quark mixing angles and the CP violating phase,
which were a main target of  the  early challenge for flavors
\cite{Pakvasa:1977in,Wilczek:1977uh}.
Since  the quark masses and mixing angles are remarkably distinguished
from the leptonic ones, that is the hierarchical structure of masses and mixing angles,  it is challenging to reproduce
observed hierarchical three CKM mixing angles and the CP violating phase 
in the $A_4$ modular symmetry
\footnote{Recently, the $S_3$ modular symmetry is also applied to the quark sector \cite{Kobayashi:2018wkl}.}.

We can easily construct quark mass matrices by using the $A_4$ modular  symmetry.
The up-quark and down-quark mass matrices have the same structure as the charged lepton mass matrix
in Ref.\cite{Kobayashi:2018scp}.
Then, parameters, apart from the modulus $\tau$,  are determined by  the  observed quark masses. The remained parameter is only  the modulus $\tau$. 
However, it is very difficult to  reproduce
observed  three CKM mixing angles
by fixing  $\tau$ since the observed mixing angles are 
considerably hierarchical angles, and moreover, precisely measured.

Therefore,  we extend the Higgs  sector in the  $A_4$ modular  symmetry by introducing the $A_4$ triplet for Higgs doublets in up-quark
and down-quark sectors, respectively. Then,  one complex parameter related with the $A_4$ tensor product appears in each quark mass matrix of 
  the up- and down-quarks.
  The model has six real parameters and two complex parameters
  in addition to the modulus $\tau$.
  It is remarked that those  quark mass matrices can predict the magnitude of the CP violation
   of the CKM mixing by inputting quark masses and three mixing angles.


The paper is organized as follows.
In section 2,  we give a brief review on the modular symmetry. 
In section 3, we present the model for quark mass matrices.
In section 4, we present  numerical results.
Section 5 is devoted to a summary.
In Appendix A, the relevant  multiplication rules of the $A_4$ group
 is presented.
In Appendix B, we show how to determine the coupling coefficients of quarks.
In Appendix C, we discuss the Higgs potential in our model.

\section{Modular group and modular forms}

The modular group $\bar\Gamma$ is the group of linear fractional transformation
$\gamma$ acting on the complex variable $\tau$, so called modulus,
belonging to the upper-half complex plane as:
\begin{equation}\label{eq:tau-SL2Z}
\tau \longrightarrow \gamma\tau= \frac{a\tau + b}{c \tau + d}\ ,~~
{\rm where}~~ a,b,c,d \in \mathbb{Z}~~ {\rm and }~~ ad-bc=1, 
~~ {\rm Im} [\tau]>0 ~ ,
\end{equation}
 which is isomorphic to  $PSL(2,\mathbb{Z})=SL(2,\mathbb{Z})/\{I,-I\}$ transformation.
This modular transformation is generated by $S$ and $T$, 
\begin{eqnarray}
S:\tau \longrightarrow -\frac{1}{\tau}\ , \qquad\qquad
T:\tau \longrightarrow \tau + 1\ ,
\end{eqnarray}
which satisfy the following algebraic relations, 
\begin{equation}
S^2 =\mathbb{I}\ , \qquad (ST)^3 =\mathbb{I}\ .
\end{equation}

 We introduce the series of groups $\Gamma(N)~ (N=1,2,3,\dots)$ defined by
 \begin{align}
 \begin{aligned}
 \Gamma(N)= \left \{ 
 \begin{pmatrix}
 a & b  \\
 c & d  
 \end{pmatrix} \in SL(2,\mathbb{Z})~ ,
 ~~
 \begin{pmatrix}
  a & b  \\
 c & d  
 \end{pmatrix} =
  \begin{pmatrix}
  1 & 0  \\
  0 & 1  
  \end{pmatrix} ~~({\rm mod} N) \right \}
 \end{aligned} .
 \end{align}
 For $N=2$, we define $\bar\Gamma(2)\equiv \Gamma(2)/\{I,-I\}$,
while, since the element $-I$ does not belong to $\Gamma(N)$,
  for $N>2$, we have $\bar\Gamma(N)= \Gamma(N)$,
  which are infinite normal subgroup of $\bar \Gamma$, called principal congruence subgroups.
   The quotient groups defined as
   $\Gamma_N\equiv \bar \Gamma/\bar \Gamma(N)$
  are  finite modular groups.
In this finite groups $\Gamma_N$,   $T^N=\mathbb{I}$  is imposed.
 The  groups $\Gamma_N$ with $N=2,3,4,5$ are isomorphic to
$S_3$, $A_4$, $S_4$ and $A_5$, respectively \cite{deAdelhartToorop:2011re}.

Modular forms of  level $N$ are 
holomorphic functions $f(\tau)$  transforming under the action of $\Gamma(N)$ as:
\begin{equation}
f(\gamma\tau)= (c\tau+d)^kf(\tau)~, ~~ \gamma \in \Gamma(N) ^.
\end{equation}
where $k$ is the so-called as the  modular weight.

Superstring theory on the torus $T^2$ or orbifold $T^2/Z_N$ has the modular symmetry \cite{Lauer:1989ax,Lerche:1989cs,Ferrara:1989qb,Cremades:2004wa,Kobayashi:2017dyu,Kobayashi:2018rad}.
Its low-energy effective field theory is described in terms of  supergravity theory,
and  string-derived supergravity theory has also the modular symmetry.
Under the modular transformation of Eq.(\ref{eq:tau-SL2Z}), chiral superfields $\phi^{(I)}$ 
transform as \cite{Ferrara:1989bc},
\begin{equation}
\phi^{(I)}\to(c\tau+d)^{-k_I}\rho^{(I)}(\gamma)\phi^{(I)},
\end{equation}
where  $-k_I$ is the modular weight and $\rho^{(I)}(\gamma)$ denotes an unitary representation matrix of $\gamma\in\Gamma(N)$.

The kinetic terms of their scalar components are written by 
\begin{equation}
\sum_I\frac{|\partial_\mu\phi^{(I)}|^2}{(-i\tau+i\bar{\tau})^{k_I}} ~,
\label{kinetic}
\end{equation}
which is invariant under the modular transformation.
Here, we use the convention that the superfield and its scalar component are denoted by the same letter.
Also, the superpotential should be invariant under the modular symmetry.
That is, the superpotential should have vanishing modular weight in global supersymmetric models, 
while the superpotential in supergravity should be invariant under the modular symmetry up to the K\"ahler 
transformation.
In the following sections, we study global supersymmetric models, e.g. minimal supersymmetric standard model 
(MSSM) and its extension with Higgs $A_4$ triplet.
Thus, the superpotential has vanishing modular weight.
The modular symmetry is broken by the vacuum expectation value of $\tau$, i.e. at the compactification scale, 
which is of order of the planck scale or slightly lower scale.

For $\Gamma_3\simeq A_4$, the dimension of the linear space 
${\cal M}_k(\Gamma_3)$ 
of modular forms of weight $k$ is $k+1$ \cite{Gunning:1962,Schoeneberg:1974,Koblitz:1984}, i.e., there are three linearly 
independent modular forms of the lowest non-trivial weight $2$.
These forms have been explicitly obtained \cite{Feruglio:2017spp} in terms of
the Dedekind eta-function $\eta(\tau)$: 
\begin{equation}
\eta(\tau) = q^{1/24} \prod_{n =1}^\infty (1-q^n)~,
\end{equation}
%
where $q = e^{2 \pi i \tau}$ and $\eta(\tau)$ is a modular form of weight~$1/2$. 
In what follows we will use the following basis of the 
$A_4$ generators  $S$ and $T$ in the triplet representation:
\begin{align}
\begin{aligned}
S=\frac{1}{3}
\begin{pmatrix}
-1 & 2 & 2 \\
2 &-1 & 2 \\
2 & 2 &-1
\end{pmatrix},
\end{aligned}
\qquad \qquad
\begin{aligned}
T=
\begin{pmatrix}
1 & 0& 0 \\
0 &\omega& 0 \\
0 & 0 & \omega^2
\end{pmatrix}, 
\end{aligned}
\label{STbase}
\end{align}
%
where $\omega=e^{i\frac{2}{3}\pi}$ .
The  modular forms of weight 2 $(Y_1(\tau),Y_2(\tau),Y_3(\tau))$ transforming
as a triplet of $A_4$ can be written in terms of 
$\eta(\tau)$ and its derivative \cite{Feruglio:2017spp}:
\begin{eqnarray} 
\label{eq:Y-A4}
Y_1(\tau) &=& \frac{i}{2\pi}\left( \frac{\eta'(\tau/3)}{\eta(\tau/3)}  +\frac{\eta'((\tau +1)/3)}{\eta((\tau+1)/3)}  
+\frac{\eta'((\tau +2)/3)}{\eta((\tau+2)/3)} - \frac{27\eta'(3\tau)}{\eta(3\tau)}  \right), \nonumber \\
Y_2(\tau) &=& \frac{-i}{\pi}\left( \frac{\eta'(\tau/3)}{\eta(\tau/3)}  +\omega^2\frac{\eta'((\tau +1)/3)}{\eta((\tau+1)/3)}  
+\omega \frac{\eta'((\tau +2)/3)}{\eta((\tau+2)/3)}  \right) , \label{Yi} \\ 
Y_3(\tau) &=& \frac{-i}{\pi}\left( \frac{\eta'(\tau/3)}{\eta(\tau/3)}  +\omega\frac{\eta'((\tau +1)/3)}{\eta((\tau+1)/3)}  
+\omega^2 \frac{\eta'((\tau +2)/3)}{\eta((\tau+2)/3)}  \right)\,.
\nonumber
\end{eqnarray}
%
The overall coefficient in Eq. (\ref{Yi}) is 
one possible choice; 
it cannot be uniquely determined.
The triplet modular forms of weight 2
have the following  $q$-expansions:
\begin{align}
Y=\begin{pmatrix}Y_1(\tau)\\Y_2(\tau)\\Y_3(\tau)\end{pmatrix}=
\begin{pmatrix}
1+12q+36q^2+12q^3+\dots \\
-6q^{1/3}(1+7q+8q^2+\dots) \\
-18q^{2/3}(1+2q+5q^2+\dots)\end{pmatrix}.
\end{align}
%
They satisfy also the constraint \cite{Feruglio:2017spp}:
\begin{align}
(Y_2(\tau))^2+2Y_1(\tau) Y_3(\tau)=0~.
\label{condition}
\end{align}

\section{Quark mass matrices in the $A_4$ triplet Higgs model}

Let us consider a $A_4$ modular invariant flavor model  for quarks.
In order to construct  models with minimal number of parameters, we introduce no flavons. There are freedoms for the assignments of irreducible representations and modular weights to quarks and Higgs doublets.
 We take similar assignments of the left-handed quarks  
  and right-handed one
 as seen in the charged lepton sector \cite{Kobayashi:2018scp}: that is,
three left-handed quark doublets are of a triplet of  $A_4$, 
 and  ($u^c, c^c, t^c$) and  ($d^c, s^c, b^c$) 
are of three different singlets 
  $\bf (1,1'',1')$ of  $A_4$, respectively. 
  For both left-handed quarks and right-handed quarks, the modular weights
  are assigned to be  $-1$, while the modular weight
  is $0$ for Higgs doublets.
Then, there appear three independent couplings in the superpotential of 
the up-quark sector and down-quark sector, respectively:

 \begin{align}
 w_u&=\alpha_u u^c H_uYQ+\beta_u c^c H_uYQ+\gamma_u t^c H_uYQ~,\label{upquark} 
 \end{align}
\begin{align}
w_d&=\alpha_d d^c H_dYQ+\beta_d s^c H_dYQ+\gamma_d b^c H_dYQ~,
\label{downquark}
\end{align}
where $Q$ is the left-handed $A_4$ triplet quarks,
and $H_q$ is the Higgs doublets.
The parameters $\alpha_q$,  $\beta_q$,  $\gamma_q$ ($q=u,d$)
are constant coefficients.
If the Higgs doublets $H_q$ are  singlet of $A_4$,
the quark mass matrices are simple form.
By using the decomposition of the $A_4$ tensor product in Appendix A, 
the supertoptential in Eqs.(\ref{upquark}) and (\ref{downquark}) gives 
the mass matrix of quarks, which is written in terms of
modular forms of   weight 2:
\begin{align}
\begin{aligned}
M_q=
\begin{pmatrix}
\alpha_q & 0 & 0 \\
0 &\beta_q & 0\\
0 & 0 &\gamma_q
\end{pmatrix}
\begin{pmatrix}
Y_1 & Y_3& Y_2 \\
Y_2 & Y_1 &  Y_3 \\
Y_3 &  Y_2&  Y_1
\end{pmatrix}_{RL},     \qquad (q=u, d)~,
\end{aligned}
\label{matrixSM}
\end{align}
where $\tau$ in the modular forms $Y_i(\tau)$ is  omitted.
Unknown couplings  $\alpha_q$,  $\beta_q$,  $\gamma_q$  can be adjusted to the  observed quark masses. The remained parameter is only   the modulus, 
$\tau$. 
The numerical study of the quark mass matrix in Eq.(\ref{matrixSM}) is
 rather easy. 
However, it is very difficult to  reproduce
observed three CKM mixing angles
by fixing one complex parameter $\tau$ because 
the CKM mixing angles are hierarchical ones and they have been  precisely measured. 


 Therefore, we enlarge the Higgs sector.
 Let us consider the Higgs doublets   to be one component of a $A_4$ triplet
 \cite{Toorop:2010ex,Degee:2012sk,Felipe:2013ie,
 	Felipe:2013vwa,Pramanick:2017wry}
  for each up-quark  and down-quark, respectively as follows:
We introduce  $A_4$ triplets Higgs  $H_u$ and $H_d$,
 which are gauge doublets, as follows: 
 \begin{align}
 \begin{aligned}
 H_u=
 \begin{pmatrix}
 H_{u1}\\
 H_{u2} \\
 H_{u3}
 \end{pmatrix} ~ , \qquad
 H_d=
 \begin{pmatrix}
 H_{d1}\\
 H_{d2} \\
 H_{d3}
 \end{pmatrix}.  
 \end{aligned}
 \end{align}

Including these $A_4$ triplet Higgs, we summarize the  assignments of representations and modular weights $-k_I$  to the 
relevant fields  in Table \ref{tb:fields}.

\begin{table}[h]
	\centering
	\begin{tabular}{|c||c|c|c|c|c|} \hline
		&$Q$&${(u^c(d^c),c^c(s^c), t^c(b^c))}$&$H_u$&$H_d$&$Y$\\  \hline\hline 
		\rule[14pt]{0pt}{0pt}
		$SU(2)$&$\bf 2$&$\bf 1$&$\bf 2$&$\bf 2$&$\bf 1$\\
		$A_4$&$\bf 3$& \bf(1,\ 1$''$,\ 1$'$)&$\bf 3$&$\bf 3$&$\bf 3$\\
		$-k_I$&$ -1$&$(-1,-1,-1)$&0&0&$k=2$ \\ \hline
	\end{tabular}
	\caption{
		The assignments of representations and
		 modular weights $-k_I$  to the MSSM fields, where 
		 Higgs sector is extented to the non-trivial
		 representation of $A_4$, $\bf 3$.}
	\label{tb:fields}
\end{table}
Now, the quark mass matrices are obtained by the tensor products
among the $A_4$ singlet right-handed quarks,  the $A_4$ triplet modular forms $Y(\tau)$,  the $A_4$ triplet Higgs  $H_q$ and the $A_4$ triplet  
left-handed quarks $Q$.
Since the tensor product of $3\otimes 3$ is decomposed  into the symmetric triplet
and the antisymmetric triplet as seen in Appendix A,  the $A_4$ invariant superpotential
in Eq.(\ref{upquark}) is expressed by introducing additional two parameters
$g_{u1}$ and $g_{u2}$ as:
\begin{eqnarray}
\begin{aligned}
& w_u=(\alpha_u u^c(1)+\beta_u c^c(1'')+\gamma_u t^c(1'))~ \otimes \\
& \left[
g_{u1}\begin{pmatrix}
2 H_{u1}Y_1-H_{u2} Y_3-H_{u3} Y_2\\
2H_{u3} Y_3-H_{u1} Y_2-H_{u2} Y_1\\
2H_{u2} Y_2-H_{u3} Y_1-H_{u1} Y_3\end{pmatrix} 
\oplus
g_{u2}\begin{pmatrix}H_{u2} Y_3-H_{u3} Y_2\\ H_{u1} Y_2-H_{u2} Y_1\\H_{u3} Y_1-H_{u1} Y_3\end{pmatrix}\right] 
\otimes
\begin{pmatrix}u\\ c\\t\end{pmatrix},
\end{aligned} \label{upsuper}
\end{eqnarray}
where the neutral component of  $H_{qi}$ is taken, and 
 the $A_4$ singlet component should be  extracted in the tensor product.
The up-quark mass matrix is given in terms of VEV's of  $ H_{ui}$,
  $v_{ui}$ in Appendix C and 
 modular forms $Y_i$($i=1,2,3$) as follows:
\begin{eqnarray}
\begin{aligned}
&M_u=\begin{pmatrix}
\alpha_u & 0 & 0\\
0& \beta_u &0\\
0 & 0 &\gamma_u
\end{pmatrix}
~ \times  \\ 
& \left [\frac{g_{u1} }{\sqrt{2}}
\begin{pmatrix}
2 v_{u1} Y_1-v_{u2} Y_3-v_{u3}  Y_2 &2v_{u2} Y_2-v_{u3} Y_1-v_{u1} Y_3
&2v_{u3} Y_3-v_{u1} Y_2-v_{u2} Y_1\\
2v_{u3} Y_3-v_{u1} Y_2-v_{u2} Y_1 &2 v_{u1} Y_1-v_{u2} Y_3-v_{u3} Y_2
&2v_{u2} Y_2-v_{u3} Y_1-v_{u1}  Y_3
\\
2v_{u2}  Y_2-v_{u3} Y_1-v_{u1} Y_3&2v_{u3} Y_3-v_{u1} Y_2-v_{u2} Y_1
&2 v_{u1} Y_1-v_{u2} Y_3-v_{u3} Y_2
\end{pmatrix}    \right .
\\ 
&+ \left . \frac{g_{u2} }{\sqrt{2}}
 \begin{pmatrix}
v_{u2} Y_3-v_{u3} Y_2& v_{u3} Y_1-v_{u1}Y_3
&v_{u1} Y_2-v_{u2} Y_1
\\v_{u1} Y_2-v_{u2} Y_1&v_{u2} Y_3-v_{u3} Y_2 & v_{u3} Y_1-v_{u1} Y_3
\\ v_{u3} Y_1-v_{u1}Y_3& v_{u1} Y_2-v_{u2} Y_1& v_{u2} Y_3-v_{u3} Y_2
\end{pmatrix} \right ] ~ ,
\end{aligned} 
\label{upmassmatrix}
\end{eqnarray}
where  $\alpha_u$, $\beta_u$, and $\gamma_u$ 
are taken to be real positive by rephasing right-handed quark fields
without loss of generality.
The down-quark mass matrix is also given by replacing $u$ with $d$ in Eq.(\ref{upmassmatrix}).

 The vacuum structure of our model
 is determined by the scalar potential  $V(H_{u},H_{d})$, 
 which is presented in Appendix C.
   Since the modular forms $Y_i$'s do not couple to the scalar potential
   due to   the modular weight of  $0$ for the Higgs doublets,
   the vacuum structure of the scalar potential  is independent of VEV of $\tau$.
   Therefore, the scalar potential is similar to the one in  MSSM.
  As discussed in the non-SUSY model with the $A_4$ triplet Higgs,
   there are  some choices of $v_q$'s to realize the vacuum 
   \cite{Toorop:2010ex,Degee:2012sk,Felipe:2013ie,
   	Felipe:2013vwa,Pramanick:2017wry}, which is the global minimum
   \footnote{Other different types of the global minima coexist and are degenerate. For example, 
   	$\langle H_d \rangle =\frac{1}{\sqrt{2}}~(v_{d}, v_{d},v_{d})$
   	and 	$\langle H_u \rangle =\frac{1}{\sqrt{2}}(v_{u}, v_{u},v_{u})$
   	lead to the global minimum.
   	Upon small variation of the parameters around this special point, one minimum
   	point becomes the global minimum while the other turns into a local one, and it is clearly possible to make either of them the global minimum
   	\cite{Degee:2012sk}.}.
   In our work, we take the simplest one   of $\langle H_q \rangle$
     in our SUSY framework as follows:
    \begin{equation}
    \langle H_u \rangle =\frac{1}{\sqrt{2}}~(v_{u1}, 0,0) ~ , \qquad
     \langle H_d \rangle =\frac{1}{\sqrt{2}}~(v_{d1}, 0,0) ~,
     \label{vq}
    \end{equation}
    in the basis of $S$ and $T$ in Eq.(\ref{STbase}).
    Here $v_{u1}$ and $v_{d1}$ are taken to be real and 
     $v_{u1}^2+v_{d1}^2=2 v_H^2$ where $v_H=174.1$GeV.
     The vacuum alignment in Eq.(\ref{vq})
      easily realizes the minimum of the scalar potential 
       by taking the condition 
       \begin{align}
       \frac{\partial V(H_u, H_d)}{\partial H_{qk}}=0 ~ ,  
       \quad (q=u,d~; ~ k=1,2,3)~,
       \end{align}
       while   the Hessian
       \begin{align}
       \frac{\partial^2  V(H_u, H_d)}{\partial  H_{qk} \partial  H_{qj}} ~ ,
       \quad (q=u,d~; ~ k,j=1,2,3) ~,
       \end{align}
       is required to have  non-negative eigenvalues, which correspond to  that
       all physical  masses being positive except for vanishing masses of 
       the Goldstone bosons  as seen   in Appendix C.
      
       
       Indeed, we have  checked  numerically 
       for $\tan\beta=v_{u1}/v_{d1}=10$ that
       the extra scalars and pseudo-scalars  could be 
        ${\cal O}(10)$TeV keeping the light  SM Higgs mass.
        This situation is  achieved due to some fine-tuning and rather large scalar self-couplings
        by taking account of the radiative corrections of
        SUSY and $\tilde m_{H_q}={\cal O}(10)$TeV,
         $B={\cal O}(10)$TeV and  $\mu={\cal O}(10)$TeV.
        However, loop corrections to the scalar masses become important
        as shown in two Higgs doublet model 	\cite{Basler:2017nzu,Krauss:2018thf}.
        Therefore, 
         such high splittings of scalar masses should be carefully examined
         in the contex of the phenomenology.
         Moreover, 
           there could be unsuppressed flavor changing neutral current (FCNC) of quarks, which was discussed in the $A_4$ triplet Higgs model \cite{Toorop:2010ex}.
          Indeed,  the study of FCNC in Kaon and B meson systems
          is important. 
          However, we do not discussed the phenomenology, which is out of scope in the present work.

        
        In our model, only  $H_{u1}$ and $H_{d1}$ have VEVs, therefore,
        it is easy to find that the couplings to the observed $125$GeV Higgs boson are expected to be proportional to quark masses.
       This situation is  understandable  since
       $H_{q1}$ do not mix with $H_{q2}$ and $H_{q3}$
        in the Higgs potential as seen in Appendix C.
       The electromagnetism is not broken: 
       a minimum of the potential satisfying 
       $\partial V/\partial H_{qk}^{\pm}=0$ 
       gives $\langle H_{qk}^{\pm} \rangle =0$.
  
     It is also noticed that the VEV in Eq.(\ref{vq}) has a residual $Z_2$ symmetry of $A_4$.
      However, this $Z_2$ symmetry of the Higgs sector is  accidental
       since an obtained   $\tau$ of our result breaks completely  $A_4$ symmetry.
     The choice of $(v_q, 0,0)$ should be considered  to reduce the number of free parameters.
     Indeed, the numerical fit of experimental data of the CKM matrix is improved by using  another alignment of $(v_q, v'_q,0)$, which has no  the $Z_2$ symmetry.


Finally, we obtain the up-quark and down-quark mass matrices:
\begin{align}
\begin{aligned}
M_q= \frac{1 }{\sqrt{2}}v_{q1}~ g_{q1}
\begin{pmatrix}
\alpha_q & 0 & 0 \\
0 &\beta_q & 0\\
0 & 0 &\gamma_q
\end{pmatrix}
\begin{pmatrix}
2Y_1 & -(1+g_q)Y_3& -(1-g_q)Y_2 \\
-(1-g_q) Y_2 & 2Y_1 &  -(1+g_q)Y_3 \\
-(1+g_q)Y_3 & -(1-g_q) Y_2& 2 Y_1
\end{pmatrix}_{RL},     ~ (q=u, d),
\end{aligned}
\label{Qmassmatrix}
\end{align}
where $g_q\equiv g_{q2}/g_{q1}~  (q=u, d)$.
 There are  six real parameters $\alpha_q$,  $\beta_q$,  $\gamma_q$ ($q=u,d$),
and the  VEV of the modulus, $ \tau $.
 In addition, we have two complex parameters $g_u$ and  $g_d$.
 It is noted that the factor $v_{q1} g_{q1}$ in front of
 the right hand side of Eq.(\ref{Qmassmatrix}) is absorbed into 
 $\alpha_q$,  $\beta_q$ and   $\gamma_q$. Thus,
  we have  six real parameters and three complex ones.
  That is to say, there are twelve free  real parameters in our mass matrices.
  It is also noticed that   $v_{q1}$ does not appear explicitly
   in our  calculations because it is absorbed in 
   $\alpha_q$,  $\beta_q$ and   $\gamma_q$.  Therefore,
   our numerical result is independent of $\tan\beta=v_{u1}/v_{d1}$.
   
 The quark mass matrix in Eq.(\ref{Qmassmatrix}) has a specific flavor structure due to the $A_4$ symmetry.  It is easily found 
 relations among matrix elements as follows:
    \begin{align}
\frac{M_q(1,1)}{M_q(2,2)}=\frac{M_q(1,2)}{M_q(2,3)}=\frac{M_q(1,3)}{M_q(2,1)}~,\qquad
\frac{M_q(2,2)}{M_q(3,3)}=\frac{M_q(2,1)}{M_q(3,2)}=\frac{M_q(2,3)}{M_q(3,1)}~.
 \end{align}
 Moreover, a constraint among $Y_1$, $Y_2$ and $Y_3$
in Eq.(\ref{condition}) provide a relation
   \begin{align}
  \frac{M_q(2,1)}{M_q(2,2)}=\frac{(g_q-1)^2}{g_q+1}\frac{M_q(3,1)}{M_q(3,2)}~ .
  \end{align}
 These relations correlate CKM mixing angles each other.
  Thus, the three  CKM mixing angles are not independent  in our quark mass matrix. 
   Indeed,  parameter region of $\tau$, $g_u$ and $g_d$
are restricted to be in rather narrow regions
in order to reproduce the three CKM mixing angles,  as seen in numerical result.
Then, the CP violating phase is predicted in the restricted region
in spite of the excess of parameters compared with observed ones. 

	
\section{Numerical results}

Let us begin with explaining how to get our prediction
of the CP violation in terms of twelve real parameters.
At first, we take a random point of  $\tau$ and  $g_u$, $g_d$,
which are scanned  in the complex plane 
by generating random numbers. The scanned ranges of 
${\rm Im } [\tau]$  is $[0.5,10]$, in which the lower-cut $0.5$ comes from the
accuracy of calculating modular functions, and  the upper-cut $10$ is enough large for estimating  $Y_i$ in practice.
On the other hand,   ${\rm Re } [\tau]$ is scanned in
the fundamental region of  $[-3/2, 3/2]$ in Eq.(\ref{Yi})
because the modular function $Y_i$  is  given in terms of   $\eta(\tau/3)$.
We also scan  in 
$|g_u| \in [0,1000]$ and $|g_d|\in [0,1000]$ while these phases are scanned
in $[-\pi,\pi]$.

Then, parameters $\alpha_q$,  $\beta_q$,  $\gamma_q$ ($q=u,d$)
are determined  by  computing functions $C^q_i(i=1-3)$ in 
Appendix B after inputting six quark masses (see Appendix B).
We use the six quark masses at the  $M_Z$ scale \cite{Antusch:2013jca}.

Finally, we can calculate three CKM mixing angles  in terms of the model parameters $\tau$, $g_u$ and  $g_d$,
while keeping the parameter sets leading to values allowed by the experimental data of the CKM mixing angles.
We continue this procedure to obtain enough points for plotting allowed region.

\begin{figure}[t!]
	\begin{minipage}[]{0.47\linewidth}
		\includegraphics[{width=\linewidth}]{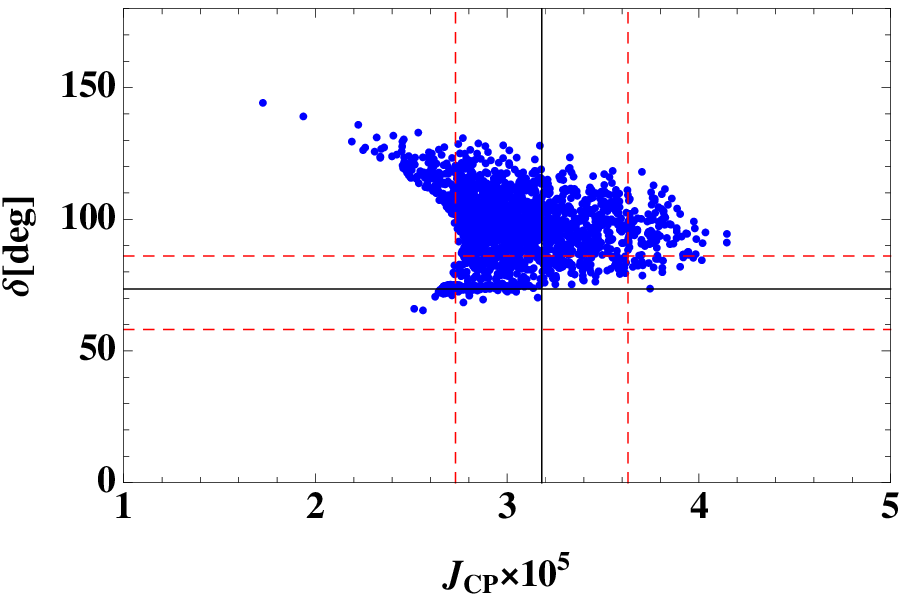}
		\caption{Prediction of the magnitude of the CP violation
			on $J_{CP}$~--~$\delta$ plane,
			where black lines denote  observed central values of 
			$J_{CP}$ and  $\delta$,  and  red dashed-lines denote
			their upper-bounds and lower-bounds of  $3\sigma$ interval.	}
	\end{minipage}
	\hspace{5mm}
	\begin{minipage}[]{0.47\linewidth}
		\vspace{-5mm}
		\includegraphics[{width=\linewidth}]{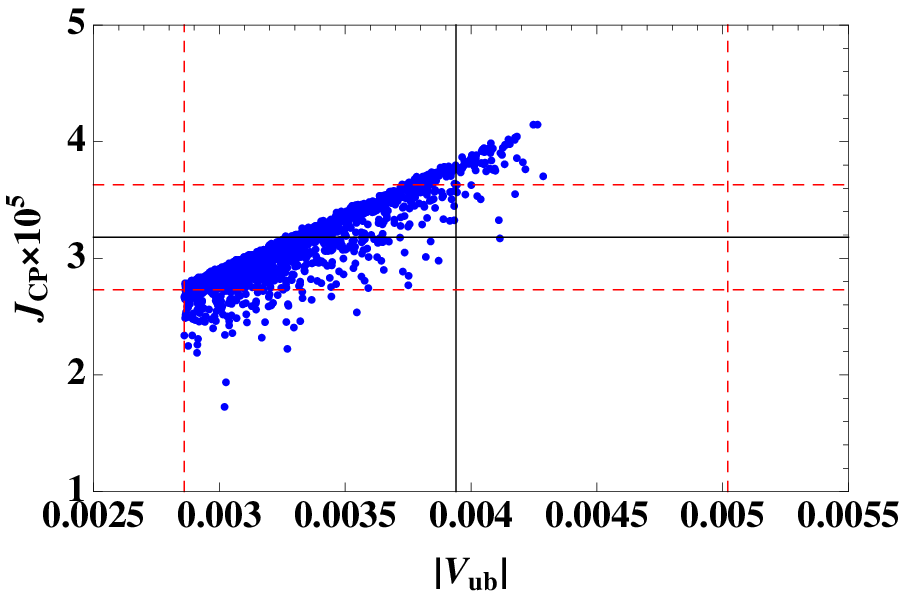}
		\caption{Predicted $J_{CP}$ versus  $|V_{ub}|$,
			where black lines denote  observed central values of 
			$|V_{ub}|$ and 	$J_{CP}$,   and  red dashed-lines denote
			their upper-bounds and lower-bounds of  $3\sigma$ interval.}
	\end{minipage}
\end{figure}

We adopt the  data of quark Yukawa couplings at the $M_Z$ scale as input
in order to constraint the model parameters
\cite{Antusch:2013jca} :
\begin{eqnarray}
&&y_d=(1.58^{+0.23}_{-0.10}) \times 10^{-5}, \quad 
y_s=(3.12^{+0.17}_{-0.16}) \times 10^{-4}, \quad ~
y_b=(1.639\pm 0.015) \times 10^{-2},  \nonumber \\
	\rule[15pt]{0pt}{1pt}
&&y_u=(7.4^{+1.5}_{-3.0}) \times 10^{-6}, \quad ~~
y_c=(3.60{\pm 0.11}) \times 10^{-3}, \quad
y_t=0.9861^{+0.0086}_{-0.0087}   ~~,
\label{yukawa}
\end{eqnarray}
which give quark masses as $m_q=y_q v_H$ with $v_H=174.1$ GeV.
We also take 
   the absolute values of CKM elements  $V_{us}$, $V_{cb}$
   and  $V_{ub}$ for input as follows \cite{Tanabashi:2018oca}:
   \begin{eqnarray}
  |V_{us}|=0.2243\pm 0.0005 ~ , \quad 
   |V_{cb}|=0.0422\pm 0.0008 ~ , \quad 
   |V_{ub}|=(3.94\pm 0.36)\times 10^{-3} ~  .
  \label{CKM}
   \end{eqnarray}
In Eqs.(\ref{yukawa}) and (\ref{CKM}), the error-bars denote interval of  $1\sigma$, and $3\sigma$  error-bars are used as input.

The obtained parameter region of $\tau$, $g_u$ and  $g_d$
 are as follows:
 \begin{eqnarray}
 &&{\rm Re}[\tau]= -(1.49-1.50) ~, \qquad\qquad\qquad {\rm Im}[\tau]= 2.01-2.02 ~ , \nonumber \\
 \rule[14pt]{0pt}{0pt}
 &&{\rm Re}[g_u]= 0.70-0.93 ~,~ \qquad\quad\qquad\qquad {\rm Im}[g_u]=\pm (0.002-0.022) ~ , 
 \nonumber \\
 \rule[14pt]{0pt}{0pt}
 &&{\rm Re}[\frac{1}{g_d}]\simeq -(0.99-1.03)\times 10^{-3} ~,
 ~ \qquad {\rm Im}[\frac{1}{g_d}]= -(0.052-0.108)~ ,
 \label{parameters}
 \end{eqnarray}
 where the modulus $\tau$ is almost fixed.
By using these values,
 we can predict the CP violation phase $\delta$ and the Jarlskog invariant $J_{CP}$~\cite{Jarlskog:1985ht}.  Those are compared with the observed values
at the electroweak scale \cite{Tanabashi:2018oca}:
\begin{eqnarray}
  \delta= (73.5^{+4.2}_{-5.1})^\circ  ~ , 
   \qquad\qquad J_{CP}=(3.18\pm 0.15)\times 10^{-5}~ .
\label{CP}
\end{eqnarray}

\begin{wrapfigure}{r}{8.2 cm}
	\vspace{0mm}
	\includegraphics[{width=\linewidth}]{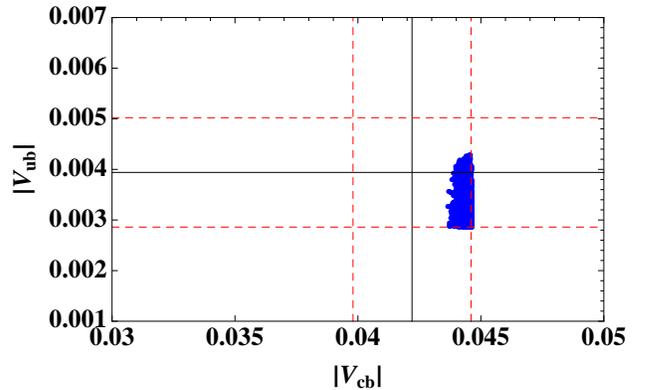}
	\caption{The allowed region on  $|V_{cb}|$--$|V_{ub}|$
		plane. Notations are same in Figs. 1 and 2.
	}
\end{wrapfigure}
Our predictions are presented in Figs.1--3.
We show the predicted  CP violating phase $\delta$
versus $J_{CP}$ in Fig.1.
Here, the observed CKM mixing elements $|V_{us}|$, $|V_{cb}|$ and $|V_{ub}|$ are input with $3\sigma$ error interval.
The predicted ranges of  $\delta$ and $J_{CP}$ is
($65^\circ$--$140^\circ$) and ($2$--$4$)$\times 10^{-5}$, respectively.
 Those  include the allowed regions of the experimental data 
 in Eq.(\ref{CP}),
 which are denoted by  red dashed-lines with $3\sigma$ error interval.
 The predicted region of  $\delta$ is still broad.
 It is remarked that $\delta$ is more restricted if
   error-bars of inputting quark masses are reduced, especially,
     the s-quark mass and the  c-quark mass
    are important to predict $\delta$.

We show the $|V_{ub}|$ dependence of predicted $J_{CP}$ in Fig.2.
Although observed $|V_{ub}|$ $[0.0028,0052]$ is input, our model does not allow the region larger than $0.0043$.
The  $|V_{ub}|$ is cut below the lower-bound of
experimental data.
The predicted $J_{CP}$ is approximately  proportional to $|V_{ub}|$. 
The upper hard cut of $J_{CP}$ is due to the maximal value of $\sin\delta=1$.

In Fig 3, we show the allowed region  on $|V_{cb}|$--$|V_{ub}|$ plane. The $|V_{cb}|$ is restricted in the very narrow range, which
is lager than $0.0436$, 
 close to the $3\sigma$ upper-bound of the observed one $0.0446$.
 This prediction provides us a crucial test of our model.

We can also discuss the ratio of  CKM matrix elements of $V_{ub}$ and $V_{cb}$,
which is  in the range of  $[0.065,0.098]$
from Fig.3.
It should be compared with the observed values \cite{Aaij:2015bfa}:
\begin{eqnarray}
\left| \frac{V_{ub}}{V_{cb}} \right | =0.083\pm 0.006~ .
\label{CKMratio}
\end{eqnarray}
Our prediction is inside of 
the observed  $3\sigma$ interval in Eq.(\ref{CKMratio}).
This measurement was given 
in the semileptonic decays of $\Lambda_b$ at LHCb.
 This prediction provides another complementary test of our model.


 Finally, 
we  show a typical set with twelve parameters as  one sample,
 which gives us successful CKM parameters as well as $J_{CP}$:
  \begin{eqnarray}
 &&\tau= -1.495+ i~ 2.011 ~ , \quad
  g_u=0.918 + i~ 0.0116 ~ ,  \quad g_d=-980- i~ 18.9~ , 
   \label{sample} \nonumber\\
 && \alpha_u/ \gamma_u= 2.496\times  10^{-5}  , \quad  \beta_u/ \gamma_u=5.995\times 10^{-3}, ~\quad
 \alpha_d/ \gamma_d= 2.855\times  10^{-3}  ,   \\
 && \beta_d/ \gamma_d=3.812\times 10^{-2}, \quad 
   \tilde\gamma_u\equiv \frac{1}{\sqrt{2}}v_ug_{u1} \gamma_u=85.85{\rm GeV}, 
   \quad \tilde\gamma_d\equiv \frac{1}{\sqrt{2}} v_dg_{d1} 
   \gamma_d=1.427{\rm GeV}.
   \nonumber 
 \end{eqnarray}
This set gives
    \begin{eqnarray}
&& |V_{us}|=0.224 ~ , \qquad \qquad
 |V_{cb}|=0.0443 ~ , \qquad \qquad
 |V_{ub}|=3.20\times 10^{-3} ~  , 
  \nonumber \\
&&  J_{CP}=2.98 \times 10^{-5}~ , \qquad \delta=74.9^\circ ~ ,
 \end{eqnarray}
which are remarkably consistent with the observed values. 
 It is noticed that 
 ratios of  $\alpha_q/ \gamma_q$ and $\beta_q/ \gamma_q$ $(q=u,d)$
 in Eq.(\ref{sample})
 correspond to the observed  quark mass hierarchy.

In conclusion, our quark mass matrix with the $A_4$ modular symmetry
can reproduce the CKM mixing matrix completely with observed quark masses.


\section{Summary}

We have discussed the quark mass matrices in the  $A_4$ modular symmetry,
where the $A_4$ triplet of Higgs doublets is introduced
 for each  up-quark and down-quark sectors, respectively.
The model has six real parameters and two complex parameters
 in addition to the modulus $\tau$.
 Then, we have constrained the model parameters  by inputting six quark masses and three CKM mixing angles at the electroweak scale.
We have  predicted the CP violation phase $\delta$ and the Jarlskog invariant $J_{CP}$.

The predicted ranges of  $\delta$ and $J_{CP}$ is
($65^\circ$--$140^\circ$) and ($2$--$4$)$\times 10^{-5}$, respectively.
Those  include the allowed regions of the experimental data. 
The absolute value of $V_{ub}$ is  smaller than $0.0043$.
The magnitude of $V_{cb}$ is lager than $0.0436$, which
is close to the $3\sigma$ upper-bound of the observed one.
Thus, our quark mass matrices with the  $A_4$  modular symmetry
can reproduce the CKM mixing matrix completely with observed quark masses.

Our mass matrices have been analyzed at the electroweak scale in this work.
The renormalization-group evolution from the GUT scale
to the electroweak scale  have been examined 
in some  textures of the quark mass matrix \cite{Xing:2015sva}.
The textures of the quark mass matrix are essentially stable against the
evolution.
We expect that the conclusions derived in this paper do not change much 
even if we consider the mass matrix   at the GUT scale.

 
 We will also  discuss the lepton mass matrices in the modular $A_4$ symmetry by introducing  the $A_4$ triplet of Higgs doublets
elsewhere.

\section*{Acknowledgments}
This research is supported by the Ministry of Science, ICT and Future Planning, Gyeongsangbuk-do and Pohang City (H.O.),
and also supported by JSPS Grants-in-Aid for Scientific Research 
15K05045 (MT).
H. O. is sincerely grateful for KIAS and all the members.


\appendix

\section*{Appendix}

\section{Multiplication rule of $A_4$ group}
\label{sec:multiplication-rule}
We take 
\begin{align}
\begin{aligned}
S=\frac{1}{3}
\begin{pmatrix}
-1 & 2 & 2 \\
2 &-1 & 2 \\
2 & 2 &-1
\end{pmatrix},
\end{aligned}
\qquad 
\begin{aligned}
T=
\begin{pmatrix}
1 & 0& 0 \\
0 &\omega& 0 \\
0 & 0 & \omega^2
\end{pmatrix}, 
\end{aligned}
\end{align}
where $\omega=e^{i\frac{2}{3}\pi}$ for a triplet.
In this base,
the multiplication rule of the $A_4$ triplet is
\begin{align}
\begin{pmatrix}
a_1\\
a_2\\
a_3
\end{pmatrix}_{\bf 3}
\otimes 
\begin{pmatrix}
b_1\\
b_2\\
b_3
\end{pmatrix}_{\bf 3}
&=\left (a_1b_1+a_2b_3+a_3b_2\right )_{\bf 1} 
\oplus \left (a_3b_3+a_1b_2+a_2b_1\right )_{{\bf 1}'} \nonumber \\
& \oplus \left (a_2b_2+a_1b_3+a_3b_1\right )_{{\bf 1}''} \nonumber \\
&\oplus \frac13
\begin{pmatrix}
2a_1b_1-a_2b_3-a_3b_2 \\
2a_3b_3-a_1b_2-a_2b_1 \\
2a_2b_2-a_1b_3-a_3b_1
\end{pmatrix}_{{\bf 3}}
\oplus \frac12
\begin{pmatrix}
a_2b_3-a_3b_2 \\
a_1b_2-a_2b_1 \\
a_3b_1-a_1b_3
\end{pmatrix}_{{\bf 3}\  } \ , \nonumber \\
\nonumber \\
{\bf 1} \otimes {\bf 1} = {\bf 1} \ , \qquad &
{\bf 1'} \otimes {\bf 1'} = {\bf 1''} \ , \qquad
{\bf 1''} \otimes {\bf 1''} = {\bf 1'} \ , \qquad
{\bf 1'} \otimes {\bf 1''} = {\bf 1} \  .
\end{align}
More details are shown in the review~\cite{Ishimori:2010au,Ishimori:2012zz}.

\section{$\alpha_q/\gamma_q$ and $\beta_q/\gamma_q$ in terms of quark masses}

The coefficients $\alpha_q$, $\beta_q$, and $\gamma_q$ in Eq.(\ref{Qmassmatrix})
are taken to be real positive without loss of generality.
These parameters are described in terms of the modulus $\tau$ and  quark masses.
The mass matrix is written as 
\begin{align}
\begin{aligned}
M_q= \frac{1}{\sqrt{2}}v_q~ g_{q1} \gamma_q
\begin{pmatrix}
\hat\alpha_q & 0 & 0 \\
0 &\hat\beta_q & 0\\
0 & 0 &1
\end{pmatrix}
\begin{pmatrix}
2Y_1 & -(1+g_q)Y_3& -(1-g_q)Y_2 \\
-(1-g_q) Y_2 & 2Y_1 &  -(1+g_q)Y_3 \\
-(1+g_q)Y_3 & -(1-g_q) Y_2& 2 Y_1
\end{pmatrix}_{RL},    
\end{aligned}
\end{align}
where $\hat{\alpha}_q\equiv\alpha_q/\gamma_q$ and  $\hat{\beta}_q\equiv\beta_q/\gamma_q$.
Then, we have three equations as:
\begin{align}
{\sum_{i=1}^3 m_{q_i}^2}
={\rm Tr}[M_q^\dag M_q]&=\tilde\gamma_q^2 (1+\hat\alpha_q^2+\hat\beta_q^2)\ C^q_{1}~,\label{eq:sum} \\
{\prod_{i=1}^3 m_{q_i}^2}
={\rm Det}[M_q^{\dag} M_q]&=
\tilde\gamma_q^6 \hat\alpha^2_q\hat\beta^2_q \ C^q_{2}~, \label{eq:prod}\\
\chi= \frac{{\rm Tr}[M_q^{\dag} M_q]^2-{\rm Tr}[(M_q^{\dag} M_q)^2]}{2}
&=\tilde\gamma^4_q (\hat\alpha_q^2+\hat\alpha^2_q\hat\beta_q^2+\hat\beta_q^2)~C^q_{3}~, \label{eq:chi}
\end{align}
where{$\chi\equiv m_{q_1}^2m_{q_2}^2+m_{q_2}^2m_{q_3}^2+m_{q_3}^2m_{q_1}^2$}
and $\tilde\gamma_q=(v_qg_{q1} \gamma_q)/\sqrt{2}$.
The coefficients $C^q_{1}$, $C^q_{2}$, and $C^q_{3}$ depend only on $Y_i$
and $g_q$, where
$Y_i$'s are determined if the value of modulus $\tau$ is fixed, and 
$g_q$ is an arbitrary complex coefficient.
Those are given explicitly as follows:
\begin{align}
\begin{aligned}
C^q_{1}=&4|Y_1|^2+|g_q-1|^2|Y_2|^2+|g_q+1|^2|Y_3|^2 ~ , \nonumber \\
C^q_{2}=&2~ {\rm Re}\left [{8}Y_1^3+(g_q-1)^3 Y_2^{{3}} -(g_q+1)^3Y_3^3+6(g_q^2-1)Y_1 Y_2 Y_3
\right ]~,
\nonumber \\
C^q_{3}=&16|Y_1|^4+|g_q-1|^2|Y_2|^4+|g_q+1|^2 |Y_3|^4+
 4 |g_q-1|^2 |Y_1 Y_2|^2 + 4|g_q+1|^2 |Y_1 Y_3|^2+|g_q^2-1|^2|Y_2 Y_3|^2  \nonumber \\
 &  +
 4~{\rm Re}\left [(g_q-1)^2(g_q^*+1)Y_1^*Y_2^2Y_3^* +2(g_q^{*2}-1)Y_1^2Y_2^*Y_3^*
 -(g_q+1)^2(g_q^*-1) Y_1^* Y_2^* Y_3^2 \right ] ~ .
\end{aligned}
\end{align}

Then, we obtain two equations
which describe $\hat\alpha$ and $\hat\beta$ as functions of quark masses, $\tau$
and $g_q$:
\begin{align}
\begin{aligned}
\frac{(1+s)(s+t)}t&=\frac{(\sum m_i^2/C^q_1)(\chi/C^q_3)}{\prod m_i^2/C^q_2}~,\quad\qquad
\frac{(1+s)^2}{s+t}&=\frac{(\sum m_i^2/C^q_1)^2}{\chi/C^q_3}~,
\end{aligned}
\end{align}
where we redefine the parameters $\hat\alpha_q^2+\hat\beta_q^2=s$ and $\hat\alpha_q^2\hat\beta_q^2=t$.
They are related as follows,
\begin{align}
\hat\alpha_q^2=\frac{s\pm\sqrt{s^2-4t}}2~,\quad\quad
\hat\beta_q^2=\frac{s\mp\sqrt{s^2-4t}}2~.
\label{alphabeta}
\end{align}

\section{Scalar potential of $A_4$ triplet Higgs}

The $A_4$ triplets Higgs, which are $\rm SU(2)$ gauge doublets, $H_u$ and $H_d$
are expressed as:
\begin{align}
\begin{aligned}
H_u=
\begin{pmatrix}
H_{u1}\\
H_{u2} \\
H_{u3}
\end{pmatrix} ~ , \qquad
H_d=
\begin{pmatrix}
H_{d1}\\
H_{d2} \\
H_{d3}
\end{pmatrix}.  
\end{aligned}
\end{align}
Since each component is  $\rm SU(2)$  doublet, it is written as:
\begin{align}
\begin{aligned}
H_{uk}=
\begin{pmatrix}
h_{uk}^+ \\
\frac{1}{\sqrt{2}}(v_{uk}+r_{uk}+ i z_{uk})
\end{pmatrix} ~ , \qquad
H_{dk}=
\begin{pmatrix}
\frac{1}{\sqrt{2}}(v_{dk}+r_{dk}+ i z_{dk}) \\
h_{dk}^-
\end{pmatrix}, 
\end{aligned}
\end{align}
where $v_{uk}$ and $v_{dk}$ are VEV's of $H_{uk}$ and $H_{dk}$,
respectively.

 The $A_4$ invariant superpotential of Higgs sector is written by
 \begin{align}
 w_H= \mu (H_{u_1} H_{d_1} +H_{u_2} H_{d_3} + H_{u_3} H_{d_2}). \label{Eq:sp} 
 \end{align}
 The scalar potential of the D-term is given as
 \begin{align}
 V_D &= \frac{g_2^2}{8}(H_{u_1}^\dag\sigma_a H_{u_1}+H_{u_2}^\dag\sigma_a H_{u_3}+H_{u_3}^\dag\sigma_a H_{u_2} + H_{d_1}^\dag\sigma_a H_{d_1}
 + H_{d_2}^\dag\sigma_a H_{d_3}+H_{d_3}^\dag \sigma_a H_{d_2})^2 \nonumber\\
 &+ \frac{g_Y^2}{8}(H_{u_1}^\dag H_{u_1}+H_{u_2}^\dag H_{u_3}+H_{u_3}^\dag H_{u_2}-H_{d_1}^\dag H_{d_1}-H_{d_2}^\dag H_{d_3}-H_{d_3}^\dag H_{d_2})^2~ ,
 \end{align}
 where $g_2$ and $g_Y$ are gauge couplings of $\rm SU(2)$ and $\rm U(1)$, respectively, and $\sigma_a$ (a=1-3) denote the Pauli matrix.
 
 On the other hand, the soft breaking term  under $A_4$ invariance is also given by 
 \begin{align}
 V_{soft}&= \tilde m_{Hu}^2 (H_{u_1}^\dag H_{u_1} +H_{u_2}^\dag H_{u_3} + H_{u_3}^\dag H_{u_2})
 + \tilde m_{Hd}^2 (H_{d_1}^\dag H_{d_1} +H_{d_2}^\dag H_{d_3} + H_{d_3}^\dag H_{d_2}) \nonumber \\
 &+ B\mu(H_{u_1} i\sigma_2 H_{d_1}+H_{u_2} i\sigma_2 H_{d_3}+H_{u_3} i\sigma_2 H_{d_2}+{\rm h.c.}). \label{Eq:sp-soft} 
 \end{align}
 
 The resulting Higgs potential is then given by:
 \begin{align}
 V(H_u, H_d)=&  m_{Hu}^2 H^\dag_{u_1} H_{u_1} + |\mu|^2 (|H_{u_2}|^2+|H_{u_3}|^2)+ 
 \tilde m_{Hu}^2(H^\dag_{u_2} H_{u_3}+H^\dag_{u_3} H_{u_2}) \nonumber \\
 & +m_{Hd}^2 H^\dag_{d_1} H_{d_1} + |\mu|^2 (|H_{d_2}|^2+|H_{d_3}|^2)+ 
 \tilde m_{Hd}^2(H^\dag_{d_2} H_{d_3}+H^\dag_{d_3} H_{d_2}) \nonumber \\
 &  +\frac{g_2^2}{8}(H_{u_1}^\dag\sigma_a H_{u_1}+H_{u_2}^\dag\sigma_a H_{u_3}+H_{u_3}^\dag\sigma_a H_{u_2} + H_{d_1}^\dag\sigma_a H_{d_1}
 + H_{d_2}^\dag\sigma_a H_{d_3}+H_{d_3}^\dag \sigma_a H_{d_2})^2\nonumber \\
 &+ \frac{g_Y^2}{8}(H_{u_1}^\dag H_{u_1}+H_{u_2}^\dag H_{u_3}+H_{u_3}^\dag H_{u_2}-H_{d_1}^\dag H_{d_1}-H_{d_2}^\dag H_{d_3}-H_{d_3}^\dag H_{d_2})^2
 \nonumber \\
 &+ B\mu(H_{u_1} i\sigma_2 H_{d_1}+H_{u_2} i\sigma_2 H_{d_3}+H_{u_3} i\sigma_2 H_{d_2}+{\rm h.c.})
 , \label{Eq:pot} 
 \end{align}
 where $m_{u}^2\equiv |\mu|^2+\tilde m_{Hu}^2$, $m_{d}^2\equiv |\mu|^2+\tilde m_{Hd}^2$.
 
 We can study the minima in the potential $V(H_u, H_d)$  of Eq.({\ref{Eq:pot})
 by taking the first derivative system
  \begin{align}
  \frac{\partial V(H_u, H_d)}{\partial H_{qk}}=0 ~ ,
  \quad (q=u,d~; ~ k=1,2,3)
  \end{align}
  where $H_{qk}$ is of the field $h_{uk}^+$, $h_{dk}^-$,
  $r_{uk}$, $z_{uk}$, $r_{dk}$ and  $z_{dk}$. Here,  the Hessian
 \begin{align}
 \frac{\partial^2  V(H_u, H_d)}{\partial  H_{qk} \partial  H_{qj}} ~ ,
 \quad (q=u,d~; ~ k,j=1,2,3)
\end{align}
is required to have  non-negative eigenvalues, which correspond to  that
all physical  masses being positive except for vanishing masses of 
  the Goldstone bosons. 
  
  Our Higgs potential analysis is same as in MSSM.
  Indeed, we have checked numerically by taking $\tan\beta=v_{u1}/v_{d1}=10$ that
 the extra scalar and pseudo-scalar masses are larger than
 in ${\cal O}(1)$TeV keeping the light  SM Higgs mass.


\end{document}